\documentclass[trackchanges]{aastex701}

\usepackage{amsmath,amssymb}
\usepackage{graphicx}
\usepackage{natbib}
\usepackage{hyperref}

\begin{document}

\title{Interpretable Analytic Formulae for GWTC-4 Binary Black Hole Population Properties via Symbolic Regression}

\author[0000-0001-8700-3455]{Chayan Chatterjee}
\affiliation{Department of Physics and Astronomy, Vanderbilt University\\ 2201 West End Avenue, Nashville, Tennessee - 37235,}
\affiliation{Data Science Institute, Vanderbilt University\\ 1400 18th Avenue South Building, Suite 2000, Nashville, Tennessee - 37212,}
\email[show]{chayan.chatterjee@vanderbilt.edu}

\begin{abstract}
Recent LIGO-Virgo-KAGRA (LVK) analyses have revealed complex structure in the binary black hole (BBH) population, including distinct features in the primary mass spectrum and nontrivial spin–mass correlations. However, the phenomenological models used to capture these features often lack analytic transparency, making it difficult to isolate robust physical laws from modeling artifacts. To address this, symbolic regression is applied to the posterior inference products of the GWTC-4 catalog, discovering compact, closed-form analytic expressions for four key population relationships: (i) the merger-rate evolution with redshift $\mathcal{R}(z)$; (ii) the mass-ratio dependence of the effective-spin distribution $\chi_{\rm eff}(q)$; (iii) the redshift evolution of the effective-spin distribution $\chi_{\rm eff}(z)$; and (iv) the conditional mass-ratio distributions associated with the $10\,M_\odot$ and $35\,M_\odot$ primary mass peaks. This framework successfully compresses both rigid and highly flexible models into differentiable phenomenological laws, dynamically recovering a low-redshift merger rate slope of $\gamma_0 = 3.18^{+0.83}_{-0.87}$ without assuming an a priori power-law form. The exact analytic derivatives provided by symbolic regression show that the $q$--$\chi_{\rm eff}$ and $z$--$\chi_{\rm eff}$ correlations are robustly driven by broadening of the posterior widths rather than shifts in the mean. Furthermore, qualitatively distinct functional forms for the mass-ratio distributions conditioned on the 10 $ M_{\odot}$ and 35 $ M_{\odot}$ primary mass peaks are identified. 
These closed-form expressions enable exact analytic gradient diagnostics and compact surrogate summaries, particularly for flexible numerical posteriors that are not otherwise available in low-dimensional analytic form. They also facilitate rapid downstream calculations for rate forecasting, formation channel comparison, and stochastic background estimation.
\end{abstract}


\section{Introduction} \label{sec:intro}

The LIGO--Virgo--KAGRA (LVK) gravitational-wave (GW) detector network \citep{LIGO, Virgo, KAGRA_1, LIGO_O4_performance, LIGO_DetChar_O4} has now observed more than $\sim$ 250 confident mergers through the fourth observing run \citep{GWTC-4_1}, transforming compact-object astrophysics from an event-by-event enterprise into a population science \citep{GWTC-4_populations}. Hierarchical Bayesian population inference applied to these events has revealed rich structure in the distributions of BBH masses, spins, and merger rates, using both parametric models (e.g., power-law-plus-peak mass spectra; \cite{PowerLawPeak}) and flexible non-parametric approaches (e.g., B-spline models; \cite{BSpline}) \citep{GWTC-4_populations}. \\

A central challenge involved in population inference is that interpretability has not scaled as quickly as catalog size. Early population studies often relied on relatively low-dimensional parametric models, such as power laws with cutoffs and Gaussian-like excesses, which were intentionally designed to encode specific astrophysical hypotheses. Those models remain extremely valuable because they connect directly to questions such as the pair-instability mass scale, mass-ratio preferences, and the spin properties expected from isolated or dynamical formation \citep{MandelDeMink2016, Rodriguez2016, Zevin2021}. However, as the data improve, increasingly flexible models have become necessary to avoid overcommitting to overly restrictive functional forms. This has motivated a broad move toward weakly parametric or nonparametric inference in the GW population literature. That flexibility comes with a cost. Posterior predictive distributions represented as splines, autoregressive processes, Gaussian-mixture descriptions, or gridded rate reconstructions can capture subtle structure, but they are not always easy to summarize in a form that is both human-interpretable and portable across analyses. \\

Symbolic regression (SR) offers a natural route between flexibility and interpretability. Rather than imposing a specific functional form a priori, it searches over analytic expressions and returns formulas that balance accuracy against complexity. \citet{WongCranmer2022} demonstrated that SR can distill flexible population models into compact equations and recover familiar phenomenological forms while also suggesting new empirical descriptions. This work builds on that program by applying SR directly across several draws of the posterior predictive distribution samples inferred by the GWTC-4.0 analysis \citep{GWTC-4_Populations_dataset}. This approach is timely because several of the most interesting current population questions are fundamentally questions about functional shape. For example, the low-redshift slope and turnover scale of the BBH merger-rate density encode information about delay times, star-formation history, and metallicity evolution \citep{Fishbach2018, MadauDickinson2014}. Likewise, correlations involving $\chi_\mathrm{eff}$ have emerged as potentially important probes of formation pathways \citep{Farr2017, Vitale2017, Callister2020}. \\

In this paper, SR is applied to four distinct population relationships--the merger rate $\mathcal{R}(z)$, the effective spin distribution as a function of the mass ratio, $\mu_{\chi_\mathrm{eff}}(q)$ and $\sigma_{\chi_\mathrm{eff}}(q)$, their redshift evolution $\mu_{\chi_\mathrm{eff}}(z)$ and $\sigma_{\chi_\mathrm{eff}}(z)$, and the mass-ratio distribution conditioned on the 10 M$_{\odot}$ and 35 M$_{\odot}$ peaks of the primary mass spectrum $p(q|\mathrm{peak})$. The goal is to construct compact analytic surrogates for a heterogeneous set of GWTC-4 posterior-predictive relations and to analyze them within a common diagnostic framework. For highly flexible models, particularly B-Spline, SR provides a genuine interpretability gain by compressing high-dimensional numerical structure into closed-form expressions, while for already parametric models, the same procedure serves primarily as a consistency check. This enables a common set of surrogate-based diagnostics, including low-redshift slopes, turnover locations and probabilities, and qualitative functional-form classifications, while clarifying which inferred features are robust across different underlying parameterizations. \\


\section{Data and Method} \label{sec:data_method}

SR is an advanced machine learning technique uniquely suited for data-driven scientific discovery. Unlike standard regression, in which a functional family is specified in advance, SR searches over combinations of elementary operators and constants to identify expressions that optimally balance descriptive accuracy against algebraic complexity. In this work, the Python library \texttt{PySR} \citep{PySR} is used, which implements an efficient evolutionary search over candidate expressions and returns a Pareto set of equations spanning this trade-off. The resulting expressions provide an interpretable representation of trends that are otherwise available only numerically through posterior summaries. \\

Our method extends the approach of \cite{WongCranmer2022} in two key ways. First, rather than fitting only a single representative curve, we perform SR independently across multiple posterior draws from the released GWTC-4.0 products. This draw-by-draw strategy propagates uncertainty from the original hierarchical inference into an ensemble of symbolic expressions, from which credible regions and distributions of surrogate-based diagnostics can be constructed. Second, because the inferred surrogates are analytic, their derivatives can be evaluated directly, enabling gradient-based diagnostics for flexible models such as B-splines, where the corresponding behavior is not naturally available as a simple low-dimensional expression and would otherwise require working with high-dimensional basis coefficients or numerical reconstructions. \\

This work operates directly on the posterior predictive population products released with the GWTC-4.0 population analysis \citep{GWTC-4_populations, GWTC-4_Populations_dataset}. For the merger-rate evolution, we consider both the flexible \textsc{BSplineIID} model \citep{Edelman2022}, which provides the main opportunity for new symbolic compression and interpretation, and the parametric \textsc{PowerLawRedshift} model \citep{Fishbach2018}, included primarily as a consistency check. Similarly, for the effective-spin correlations with mass ratio and redshift, we analyze both the flexible \textsc{Spline} and simpler \textsc{Linear} parametrizations of the conditional spin distribution \citep{Talbot2017, TalbotThrane2018}, with the linear case serving mainly to test whether the SR procedure reproduces already interpretable native trends. For the mass-ratio distributions, the \textsc{Extended Broken Powerlaw + 2 Peaks} model \citep{TalbotThrane2018, GWTC3pop} with separate $p(q)$ distributions conditioned on the low- and high-mass peaks is used. The data are stored as HDF5 posterior grids of shape $(N_{\mathrm{draws}}, N_{\mathrm{grid}})$, with $N_{\mathrm{draws}}$ ranging from $\sim 150$ to $\sim 4400$ depending on the model. For each analysis, \texttt{PySR} is configured with binary operators $\{+, -, \times, /\}$ and unary operators $\{\exp, \log, \sqrt{\cdot}, \tanh, |\cdot|\}$. The search uses 300--1000 iterations, 50--80 populations, a maximum expression size of 28--45 nodes, and a parsimony penalty of $10^{-4}$--$10^{-6}$. Mean squared error is adopted as the loss, evaluated in $\log_{10}$-space for the rate and mass-ratio distributions to better capture their dynamic range. A 70/30 train-test split is used to guard against overfitting. PySR is run independently on $N=200$ posterior draws for each target relation, yielding an ensemble of best-fit symbolic expressions. From these expressions, pointwise credible intervals and the distributions of derived quantities are obtained. \\

\section{Results} \label{sec:results}

\subsection{BBH Merger Rate Evolution} \label{sec:res_Rz}

SR is applied to the BBH comoving merger-rate evolution, 
$\mathcal{R}(z)$, inferred from GWTC-4 using both the parametric \textsc{PowerLawRedshift} model \citep{Fishbach2018} and the more flexible \textsc{BSplineIID} model \citep{Edelman2022}. The latter provides the main opportunity for symbolic compression, since its inferred rate evolution is represented by a higher-dimensional and less directly interpretable reconstruction. The parametric \textsc{PowerLawRedshift} case is analyzed in parallel primarily as a consistency check, allowing both model families to be compared within the same surrogate-based diagnostic framework. All symbolic regression analyses in this paper are performed using the \texttt{PySR} software package \citep{PySR}. To physically interpret the resulting library of analytic expressions, their morphological features are quantified using derivative-based diagnostics. The steepness of the low-redshift rise is characterized using a model-independent logarithmic slope, defined as:

\begin{equation} \label{eq:gamma0}
    \gamma_0 \equiv \left\langle \frac{d\ln\mathcal{R}}{d\ln(1{+}z)} \right\rangle_{z \in [0.1,\, 0.3]},
\end{equation}

This metric measures the effective power-law index of the rate evolution at low redshift without assuming a global power-law functional form a priori. The probability of a high-redshift turnover is also assessed by calculating the fraction of posterior symbolic expressions where the derivative is negative ($d\mathcal{R}/dz < 0$) at various reference redshifts ($z_\star$). Finally, the global shape of each symbolic expression is catagorized using the automated criteria detailed in Table~\ref{Symbolic_classifications}. By evaluating the roots and signs of the first and second derivatives, this scheme maps the raw mathematical equations generated by the SR algorithm into physically meaningful phenomenological classes (e.g., distinguishing between a continuously growing rate and one that peaks and turns over). \\

As shown in Figure~\ref{fig:Rz_overlay}, the symbolic surrogates successfully reproduce the published GWTC-4 trends for both models while distilling them into closed-form equations. For the \textsc{PowerLawRedshift} model, SR reproduces the expected monotonic rate evolution with a compact surrogate, shown here as a benchmark:


\begin{equation} \label{eq:Rz_PL}
    \log_{10}\mathcal{R}(z) = \sqrt{z + \left[\left(-0.28z + 2.63\right)z + 1.66\right]} - 0.08.
\end{equation}

Analyzing the 200 individual posterior draws reveals an overwhelmingly monotonic population: 194 draws are classified as \texttt{monotonic\_powerlaw\_like}, 5 as \texttt{broken\_rise}, and 1 as \texttt{complex}. The extracted low-redshift slope is $\gamma_0 = 3.18^{+0.83}_{-0.87}$ (90\% CI). This value is robust across multiple tested low-$z$ windows (e.g., $[0.05, 0.2]$ and $[0.2, 0.4]$). Consistent with the model's structural constraints, there is zero posterior symbolic support for a declining merger rate at any tested redshift between $z = 0.5$ and $1.25$, yielding a boundary-limited peak at $z_{\rm peak} = 1.9$. By contrast, the \textsc{BSplineIID} model admits substantially more diverse and flexible functional forms. The median symbolic expression is:

\begin{equation} \label{eq:Rz_BS}
    \log_{10}\mathcal{R}(z) = \sqrt{\frac{\sqrt{z}}{z^2 - 1.53z + 1.02} + z + 1.61}.
\end{equation}

\begin{figure*}[t]
    \centering
    \includegraphics[width=\textwidth]{figures/fig1_Rz_overlay.png}
    \caption{BBH comoving merger rate $\mathcal{R}(z)$ as a function of redshift. Shaded bands show the 90\% credible intervals from the GWTC-4 \textsc{PowerLawRedshift} (blue) and \textsc{BSplineIID} (green) models. Dashed lines show the corresponding PySR symbolic regression median fits (orange), with PySR 90\% credible bands derived from 200 draw-by-draw fits. The symbolic expressions faithfully capture both the median and the uncertainty structure of the GWTC-4 posteriors.}
    \label{fig:Rz_overlay}
\end{figure*}

\begin{figure}[t]
    \centering
    \includegraphics[width=\columnwidth]{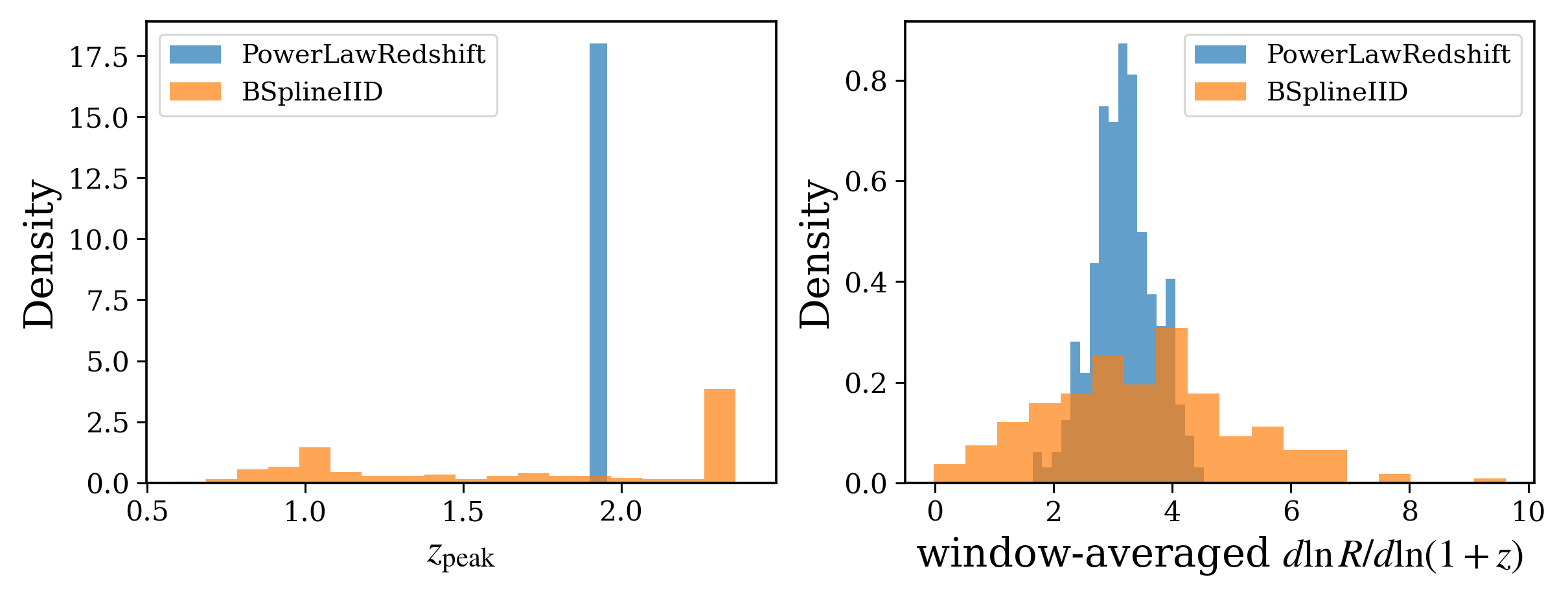}
    \caption{Distributions of the peak redshift $z_\mathrm{peak}$ (left) and the low-$z$ logarithmic slope $\gamma_0$ (right), extracted from 200 draw-by-draw PySR fits for the \textsc{PowerLawRedshift} (blue) and \textsc{BSplineIID} (orange) models. Because the \textsc{PowerLawRedshift} symbolic surrogates are overwhelmingly monotonic, they lack a true mathematical turnover. Consequently, their $z_\mathrm{peak}$ values pile up at the upper boundary of the evaluation grid ($z = 1.9$). By contrast, the more flexible \textsc{BSplineIID} model yields a broad distribution of peak redshifts yielding a statistical ensemble median to $z = 1.75$. Despite these differences in high-redshift behavior, both models strongly agree on the steep low-redshift growth rate, yielding $\gamma_0 \sim 3$--$3.5$.}
    \label{fig:Rz_diagnostics}
\end{figure}

This median relation is distinctly peaked, turning over at $z_{\rm peak} \simeq 1.01$. The posterior symbolic family is highly heterogeneous: out of 200 draws, 125 are classified as \texttt{peaked}, 66 as \texttt{complex}, 7 as \texttt{broken\_rise}, and 1 each as \texttt{monotonic\_powerlaw\_like} and \texttt{plateau\_like}. The inferred turnover is weak at low redshift but becomes moderately favored beyond $z \sim 1$, with the probability of a declining rate rising from $\mathrm{Pr}(d\mathcal{R}/dz < 0) = 0.01$ at $z = 0.5$ to $0.63$ at $z = 1.25$. The overall peak redshift across the ensemble is $z_\mathrm{peak} = 1.75^{+0.61}_{-0.92}$ (90\% CI). Despite this high-redshift flexibility, the extracted low-redshift slope remains highly consistent with the power-law model, yielding $\gamma_{0} = 3.47^{+2.89}_{-2.49}$ (90\% CI). \\

Both models independently agree on a steep low-$z$ rise consistent with $\mathcal{R}(z) \propto (1{+}z)^{\sim 3}$ (Figure~\ref{fig:Rz_diagnostics}, right panel). However, the left panel reveals a stark contrast in their implied high-redshift evolution. For the \textsc{PowerLawRedshift} model, the $z_\mathrm{peak}$ distribution is concentrated entirely in a single bin at the evaluation boundary ($z = 1.9$). This artifact arises because the model's posterior is overwhelmingly dominated by monotonically rising functions; lacking a genuine mathematical turnover, the maximum merger rate is trivially found at the upper edge of the analyzed redshift grid. In contrast, the \textsc{BSplineIID} model yields a broad, well-resolved distribution of peak redshifts centered near $z \sim 1.75$. Because the flexible spline parametrization readily admits mathematically peaked functions, this spread represents a genuine physical uncertainty regarding the exact location of the turnover rather than a boundary constraint.

\begin{table*}
\centering
\begin{tabular}{lll}
\hline
\textbf{Label} & \textbf{Operational definition} & \textbf{Interpretation} \\
\hline
\texttt{peaked} & Interior maximum of $R_{\rm SR}(z)$ & Turnover within fitted range \\
\texttt{monotonic\_powerlaw\_like} & $\frac{dR_{\rm SR}}{dz} \ge 0$ throughout, without clear flattening & Continued monotonic rise \\
\texttt{saturating} & Monotonic rise with strongly reduced high-$z$ slope & Rise toward plateau \\
\texttt{plateau\_like} & Extended interval with near-zero slope & Approximately flat evolution \\
\texttt{broken\_rise} & Monotonic rise with curvature change(s) & Bent or broken-slope rise \\
\texttt{monotonic\_decreasing} & $\frac{dR_{\rm SR}}{dz} \le 0$ throughout & Continued decline \\
\texttt{complex} & No simple class matched & Irregular or mixed behavior \\
\hline
\end{tabular}
\caption{Functional-form classification of symbolic merger-rate relations. Labels are assigned based on the numerical behavior of the symbolic prediction $R_{\rm SR}(z)$ evaluated on the GWTC-4 redshift grid, using first- and second-derivative diagnostics.}
\label{Symbolic_classifications}
\end{table*}

\subsection{Effective Spin vs.\ Mass Ratio} \label{sec:res_chieff_q}

\begin{figure*}[t]
    \centering
    \includegraphics[width=\textwidth]{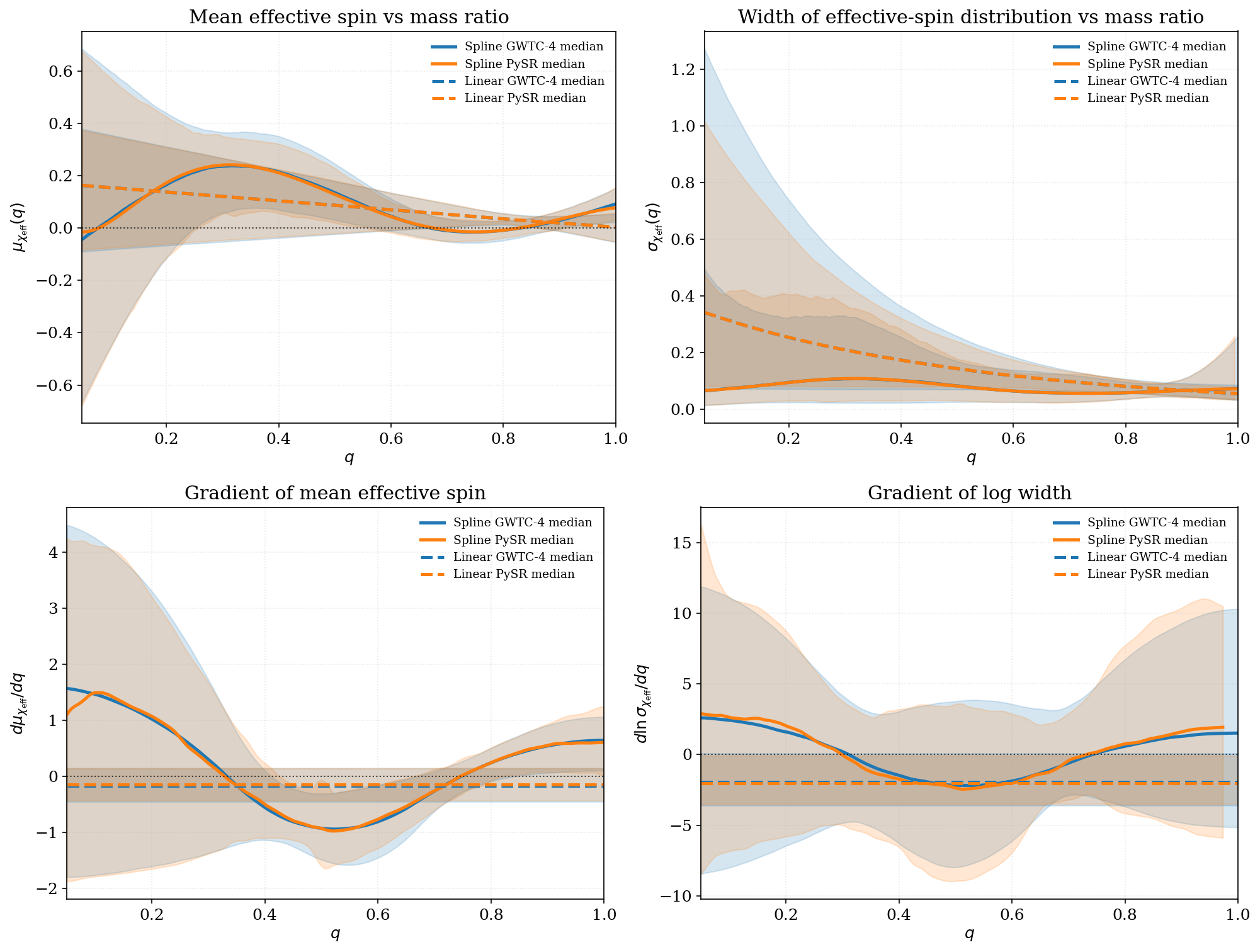}
    \caption{Effective spin distribution parameters as a function of mass ratio $q$. \textit{Top left:} Mean effective spin $\mu_{\chi_\mathrm{eff}}(q)$ for both the \textsc{Spline} (blue solid) and \textsc{Linear} (orange dashed) GWTC-4 models, with PySR symbolic fits overlaid. \textit{Top right:} Width $\sigma_{\chi_\mathrm{eff}}(q)$ of the spin distribution. \textit{Bottom left:} Analytic gradient $d\mu_{\chi_\mathrm{eff}}/dq$ computed from the PySR symbolic expressions. \textit{Bottom right:} Gradient $d\ln\sigma_{\chi_\mathrm{eff}}/dq$. Shaded bands show 90\% credible intervals. The gradients shown here are surrogate-based diagnostics constructed from the symbolic fits.}
    \label{fig:chieff_q}
\end{figure*}

Next, the GWTC-4 $q$--$\chi_{\rm eff}$ relation is analyzed using both the published \textsc{Linear} and \textsc{Spline} models \citep{Talbot2017, TalbotThrane2018, CallisterFarr2024}. Here the main added value of SR lies in the \textsc{Spline} case, where it provides a compact surrogate for a nontrivial inferred shape, while the \textsc{Linear} case serves primarily to verify that the SR pipeline faithfully recovers trends that are already transparent in the native parametrization. SR is applied independently to the 1D distributions for the mean, $\mu_{\chi_{\rm eff}}(q)$, and the logarithmic width, $\log \sigma_{\chi_{\rm eff}}(q)$. By fitting both the posterior median relations and an ensemble of 200 posterior draws, analytic surrogates are extracted that reproduce the published trends. Because SR yields closed-form mathematical expressions, these local gradients (e.g., $d\mu/dq$ evaluated at a reference $q=0.6$) are computed exactly via numerical differentiation. Evaluating these exact derivatives across the 200 draw-by-draw fits allows us to construct full posterior distributions for the slopes, directly capturing the statistical uncertainty of the morphological features. Fig.~\ref{fig:chieff_q} visualizes these reconstructed trends alongside the GWTC-4 constraints. \\

The results show that for the $q$--$\chi_{\rm eff}$ correlation, the inferred mean behavior is highly sensitive to modeling assumptions. For the \textsc{Linear} model, SR reproduces the published parametric trend:

\begin{equation}
\mu_{\chi_{\rm eff}}(q) \simeq 0.17(1 - q)
\end{equation}

This gives a weak, approximately constant gradient across the mass-ratio range. Across the posterior ensemble, the median derivative is $d\mu/dq|_{q=0.6} = -0.17^{+0.28}_{-0.22}$. Because this interval crosses zero, even the sign of the mean shift is not entirely secure under a linear prior. \\

By contrast, the more flexible \textsc{Spline} model favors a genuinely curved, non-monotonic relation, visible as a distinct dip in the median trace. A compact symbolic surrogate capturing this structure is
\begin{equation}
\mu_{\chi_\mathrm{eff}}(q)
=0.84\left(\left|\frac{5.23\,q(1-q)^2-0.24}{q(1-q)+0.82}\right|+0.05\right)^2-0.02.
\end{equation}
Unlike the \textsc{Linear} model, this surrogate is not characterized by a single constant slope. Instead, it exhibits positive window-averaged slopes near both boundaries, while developing a pronounced negative gradient at intermediate mass ratios. Defining the low-$q$ and high-$q$ slope windows as $q\in[0.05,0.3]$ and $q\in[0.7,0.9]$, respectively, the median surrogate yields average slopes of $+1.04$ at low $q$ and $+0.33$ at high $q$, together with a steep local derivative $d\mu/dq|_{q=0.6}=-0.75$. Across the posterior ensemble, the median derivative at $q=0.6$ is $d\mu/dq|_{q=0.6}=-0.80^{+0.58}_{-0.67}$ (90\% CI), while the low-$q$ and high-$q$ windows still favor positive average slopes. Thus, the \textsc{Spline} model supports a localized intermediate-$q$ downturn rather than a simple monotonic trend, making clear that the inferred mean relation is strongly model-dependent. \\

It must be noted that the disagreement near $q\to 0$ of the median \textsc{Spline} fit of $d\mu_{\chi_{\text{eff}}}/dq$, (bottom-left panel of Fig.~\ref{fig:chieff_q}) with the GWTC-4 curve, is a boundary effect of the symbolic surrogate rather than a robust physical feature. Small differences between the GWTC-4 curve and the SR median fit are amplified by differentiation at the edge of the fitted domain. \\

Unlike the mean relation, the width of the effective-spin distribution shows a comparatively robust trend across both model families, indicating that the inferred $q$--$\chi_{\rm eff}$ correlation is driven more securely by changes in the dispersion than by shifts in the mean. This is also the clearest visual feature of Fig.~\ref{fig:chieff_q}: in both models, the credible intervals narrow visibly as $q \to 1$, funneling nearly equal-mass binaries toward a much tighter range of effective spins centered near zero. For $\log \sigma_{\chi_{\rm eff}}(q)$, both model families therefore favor negative gradients over much of the mass-ratio range.  \\

For the \textsc{Linear} model, SR recovers a simple linear decline, reproducing the published trend,
\begin{equation}
\log \sigma_{\chi_{\rm eff}}(q) \simeq -1.92q - 0.99,
\end{equation}
corresponding to an approximately constant gradient across the full mass-ratio range. The median surrogate therefore yields $d\ln\sigma/dq|_{q=0.6} \simeq -1.92$ fully consistent with the expected monotonic narrowing toward equal masses. Across the posterior ensemble, the median derivative at $q=0.6$ is $d\ln\sigma/dq|_{q=0.6}
= -1.94^{+1.76}_{-1.30}$ (90\%$~\mathrm{CI}$), yielding a high posterior probability of narrowing toward equal masses, $\mathrm{Pr}[\mathrm{narrows}] = 0.98.$ \\

For the more flexible \textsc{Spline} model, a compact symbolic surrogate can reproduce the same overall narrowing trend while allowing substantially more local structure. Using the higher-fidelity surrogate adopted for the median fit,
\begin{align}
\log \sigma_{\chi_\mathrm{eff}}(q)
&= 50.5\Big[\big((1-q)^2 - 0.19\big)q(1-q)(\sqrt{q}+0.16)\,q(1-q) - 0.05\Big],
\end{align}
the median symbolic fit gives $d\ln\sigma/dq|_{q=0.6} \simeq -2.02$, again indicating narrowing toward equal masses. Across the posterior ensemble, the median derivative at $q=0.6$ is $d\ln\sigma/dq|_{q=0.6} = -1.83^{+4.95}_{-3.38} $  (90\%$~\mathrm{CI}$), with a still-robust narrowing probability of $\mathrm{Pr}[\mathrm{narrows}] = 0.81$. \\

These symbolic results indicate that the mean effect, while present, is weak and highly sensitive to parametrization. Conversely, the narrowing of the effective-spin distribution for symmetric binaries survives the transition from rigid to flexible priors, visibly constraining highly symmetric systems to a much tighter range of spin configurations than their unequal-mass counterparts. More broadly, this SR framework provides an interpretable bridge between flexible population inference and compact phenomenology: it explicitly identifies which aspects of an inferred multidimensional correlation are robust across modeling choices, and securely reduces a complex population trend into a simple, physically meaningful effective law. The gradient-based diagnostics from the surrogates provide direct constraints on how spin properties vary with mass ratio that can be compared to formation channel predictions \citep{Callister2020, q_chi_eff_Vijaykumar}. \\

\subsection{Effective Spin vs.\ Redshift} \label{sec:res_chieff_z}

\begin{figure*}[t]
    \centering
    \includegraphics[width=\textwidth]{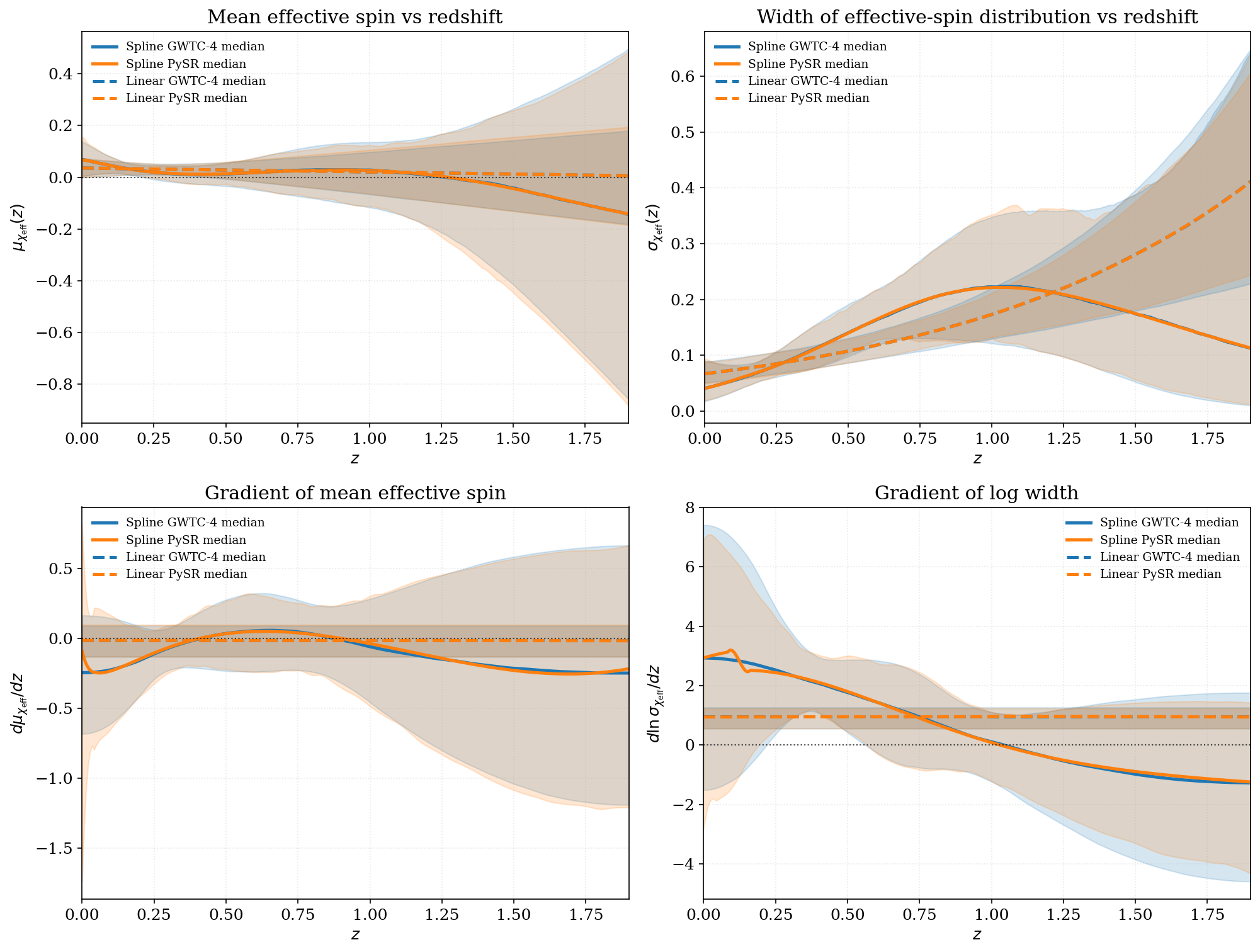}
    \caption{Effective spin distribution parameters as a function of redshift $z$. \textit{Top left:} Mean effective spin $\mu_{\chi_\mathrm{eff}}(z)$ for the \textsc{Spline} and \textsc{Linear} GWTC-4 models with PySR overlays. \textit{Top right:} Width $\sigma_{\chi_\mathrm{eff}}(z)$. \textit{Bottom left:} Analytic gradient $d\mu_{\chi_\mathrm{eff}}/dz$. \textit{Bottom right:} Gradient $d\ln\sigma_{\chi_\mathrm{eff}}/dz$.}
    \label{fig:chieff_z}
\end{figure*}

SR is applied to the redshift--effective spin ($z$--$\chi_{\rm eff}$) inference results, independently evaluating the \textsc{Linear} and \textsc{Spline} models \citep{Talbot2017, TalbotThrane2018} published in GWTC-4. As above, the main novelty is in the flexible \textsc{Spline} reconstruction, whereas the \textsc{Linear} model is used mainly as a benchmark for cross-model consistency. The joint distribution is decomposed again into 1D symbolic fits for the mean, $\mu_{\chi_{\rm eff}}(z)$, and the logarithmic width, $\log \sigma_{\chi_{\rm eff}}(z)$. These functions are evaluated across a defined low-redshift window ($z \in [0.0, 0.4]$) and a high-redshift window ($z \in [1.0, 1.9]$). Crucially, because SR extracts exact closed-form analytic expressions, we compute exact symbolic derivatives (e.g., $d\ln\sigma/dz$) for every individual fit across the 200 posterior draws. Statistical confidences, such as the probability of broadening ($\mathrm{Pr}[\text{broadens}]$), are then computed via directly counting the exact fraction of draws that exhibit a positive derivative within the relevant redshift window. Figure~\ref{fig:chieff_z} visualizes these reconstructed trends. The inferred mean relation $\mu_{\chi_{\rm eff}}(z)$ is found to be highly sensitive to modeling assumptions (Figure~\ref{fig:chieff_z}, left panel). For the \textsc{Linear} model, SR naturally extracts a shallow, monotonically declining surrogate:

\begin{equation}
    \mu_{\chi_{\rm eff}}(z) \simeq 0.05 - 0.02z
\end{equation}

The corresponding gradient curves are effectively flat and pinned near zero, yielding a median slope of $d\mu/dz \approx -0.02$. While this slight decline is uniform across all posterior draws, its absolute magnitude is physically marginal. Conversely, the more flexible \textsc{Spline} model captures a genuinely non-monotonic evolution. An optimal low-complexity \textsc{Spline} surrogate isolates this structure using a simple quadratic form:

\begin{equation}
    \mu_{\chi_{\rm eff}}(z) \simeq -0.02 z^2
\end{equation}

This relation yields a median zero-crossing at $z \approx 1.28$. However, the Spline gradient curves reveal immense posterior uncertainty, and the evolutionary direction is highly inconsistent across individual draws—with only a 75\% probability of a negative gradient in the low-$z$ window. Consequently, the structural complexity of the mean evolution is not a robustly informed feature of the data, but rather an artifact of prior flexibility. \\

In stark contrast to the mean, the redshift evolution of the distribution's width is a robust physical feature (Fig.~\ref{fig:chieff_z}, right panel). Both models favor a substantial broadening of the effective-spin distribution as redshift increases in the local universe. For the \textsc{Linear} model, SR identifies a strong, uniform monotonic broadening, cleanly represented by linear log-width surrogates:

\begin{equation}
    \log \sigma_{\chi_{\rm eff}}(z) \simeq 0.96z - 2.7
\end{equation}

The posterior gradient curves confidently exclude zero, yielding a median $d\ln\sigma/dz \approx 0.96$ and an overwhelming probability of monotonic broadening equal to unity ($\mathrm{Pr}[\text{broadens}] = 1.0$). The \textsc{Spline} model permits more complex dispersion dynamics, but SR firmly captures an initial low-redshift broadening phase. The optimal symbolic surrogate mathematically encodes this transition using an absolute-value term ($|z-1|$) to serve as a functional hinge at $z=1$:

\begin{equation}
\log\sigma_{\chi_\mathrm{eff}}(z) \simeq 0.15\,|z{-}1|\big( 1.61z^2 - 0.23z - 1.69 \big) - 1.45(z{-}1)^2 - 1.50
\end{equation}

This expression elegantly distills the Spline model's complex median behavior. By expanding the polynomial for local redshifts ($z < 1$), the equation yields a steep, continuous broadening within the low-$z$ window (average gradient $\approx +2.60$). As evolution crosses $z=1$, the absolute-value hinge smoothly inverts the polynomial contribution, transitioning the distribution into a high-redshift plateau and gradual narrowing phase (average gradient $\approx -0.75$). The reconstructed gradient curves vividly capture this two-phase behavior, cresting sharply before dipping at higher redshifts. Across the \textsc{Spline} posterior ensemble, we find a 93.4\% probability ($\mathrm{Pr} = 0.93$) that the gradient curves remain positive in the low-$z$ regime, independently confirming the growth trend seen in the \textsc{Linear} model. \\

The sharp feature in the $d\ln\sigma/dz$ SR median gradient near $z=0$ in Fig.~\ref{fig:chieff_z} arises from boundary sensitivity in the differentiated symbolic surrogate, and should be regarded as a fitting artifact. The robust conclusion is the broader low-$z$ preference for positive $d\ln\sigma/dz$, not the exact gradient shape at the boundary. \\

Astrophysically, this robust low-redshift broadening suggests a redshift-dependent evolution in the dominant BBH formation channels \citep{Zevin2021}. A primary contribution from isolated binary evolution at low redshift would yield a narrow, positively aligned spin distribution \citep{MandelDeMink2016, Bavera2022}. However, an increasing relative contribution from dynamical assembly, which yields isotropic, widely distributed spins \citep{Rodriguez2016, Farr2017}, at higher redshifts would manifest precisely as this observed broadening in $\sigma_{\chi_{\rm eff}}$ \citep{z_chi_eff_Vijaykumar}. Standard parametric models often struggle to isolate whether a shifting population is driven by its mean or its variance. By compressing posterior distributions into exact, differentiable mathematical laws, SR quantitatively proves—via explicitly extracted gradient distributions—that the apparent complexity in the population's redshift evolution is driven fundamentally by dispersion growth, rather than a macroscopic shift in the mean. \\

\subsection{Mass-Ratio Distribution by Mass Peak} \label{sec:res_pq}

\begin{figure*}[t]
    \centering
    \includegraphics[width=\textwidth]{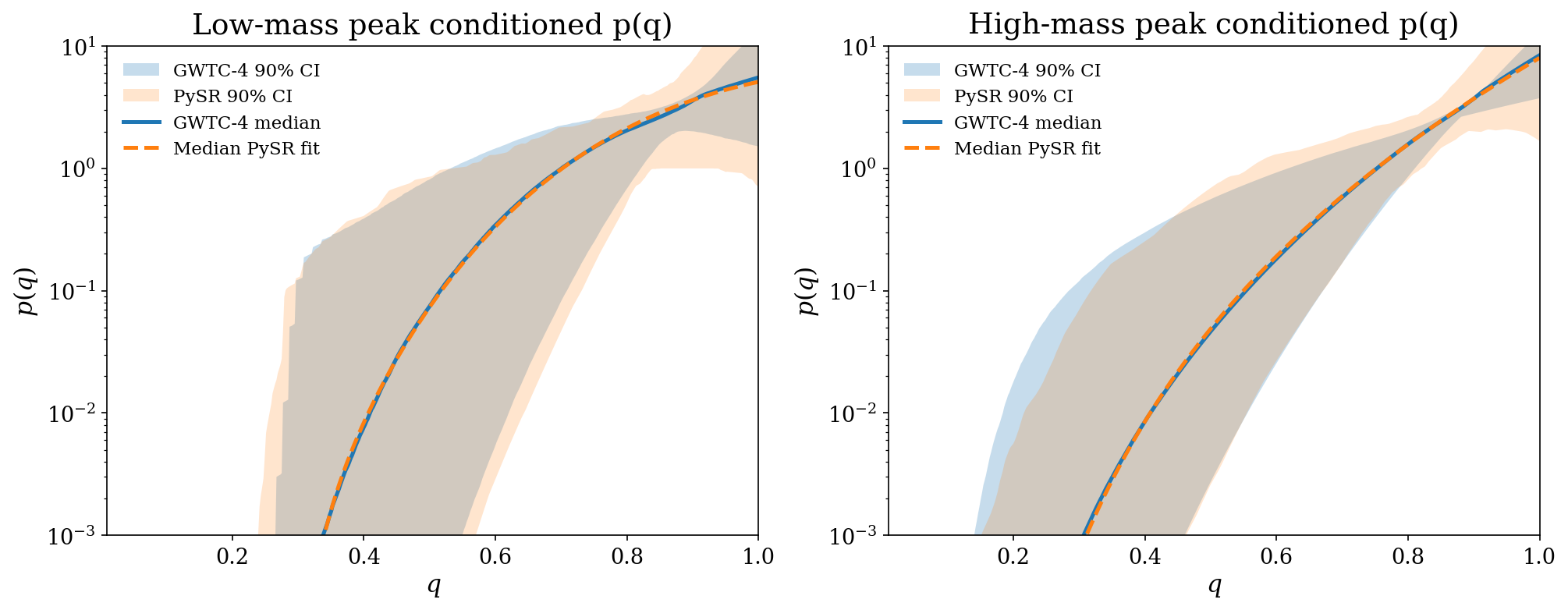}
    \caption{Conditional mass-ratio distributions $p(q)$ for the low-mass peak (\textit{left}) and high-mass peak (\textit{right}) of the BBH primary mass distribution. Blue shaded bands show the GWTC-4 90\% credible intervals, orange bands show the PySR 90\% CI from draw-by-draw fits, and the solid (GWTC-4) and dashed (PySR) lines show the medians. The low-mass peak is well-described by a simpler expression (Equation~\ref{eq:pq_low}), while the high-mass peak requires a more complex functional form (Equation~\ref{eq:pq_high}).}
    \label{fig:pq}
\end{figure*}

In GWTC-4, the parameterized \textsc{Extended Broken Power Law + 2 Peaks} model \citep{TalbotThrane2018, GWTC3pop} was used to explore whether the distribution of mass ratios of BBHs are dependent on the primary mass, specifically the low-mass ($m_1 \sim 10\,M_\odot$) and high-mass ($m_1 \sim 35\,M_\odot$) peaks of the BBH primary mass spectrum. While flexible phenomenological models like the \textsc{Extended Broken Power Law + 2 Peaks} are powerful at capturing complex structures through multidimensional parameter mixtures, they lack unified, easily manipulable analytic forms. Here, SR is applied to compress the posterior distributions into exact, analytical scaling laws. Further, by evaluating these distributions across an ensemble of 200 posterior draws, a direct, mathematically rigorous comparison of the inferred shapes across the two dominant mass components is made in this study. The inferred analytic forms can readily be integrated into binary evolution codes or dynamical encounter rates for population studies. Fig.~\ref{fig:pq} visualizes these reconstructed distributions. \\

For the low-mass component ($m_1 \sim 10\,M_\odot$), SR identifies an exceptionally accurate compact surrogate that captures the underlying complexity of the distribution:

\begin{equation} \label{eq:pq_low}
\log p(q|\mathrm{low}) \simeq 15.2\ln q + 104 \exp\big[{-3326\, e^{-39.4\, q}}\big] - 12.6\, q - 89.8
\end{equation}

The insight provided by this symbolic expression is the explicit mathematical nature of the low-$q$ deviation. The double-exponential term perfectly isolates a sharp step-function behavior, mathematically shutting off the probability density at unequal mass ratios ($q \le 0.2$). Crucially, this sharp cutoff is not an independent astrophysical discovery, but rather the exact manifestation of an indirect LVK modeling constraint. Without any explicit physical intuition or prior knowledge of the $m_{\rm min}$ parameter boundary, the SR algorithm successfully recovered this hard constraint directly from the structure of the posterior predictive grids. While this flexible functional form technically permits an interior maximum, evaluating exact mathematical diagnostics across the full SR posterior ensemble reveals that the data does not reliably support a dominant sub-equal-mass preference. Although 22\% of the symbolic fits exhibit a local interior peak, 84\% of the posterior draws ultimately reach their absolute global maximum at equal mass, yielding a posterior-median peak location pinned tightly at $q_{\rm peak} = 1.0$. \\



By contrast, the high-mass peak ($m_1 \sim 35\,M_\odot$) is much more cleanly consistent with standard equal-mass pairing, lacking the extreme low-$q$ cutoff features of its low-mass counterpart. The optimal low-complexity surrogate takes the form:

\begin{equation} \label{eq:pq_high}
\log p(q|\mathrm{high}) \simeq 7.30\ln q - \exp\Big[1.22\, e^{\sqrt{q}(9.78 - 100.2\, q)}\Big] - [\ln(q + 0.24)\ln q]^2 + 3.08
\end{equation}

This high-mass surrogate decisively favors a peak at $q_{\rm peak} \simeq 1.0$. Across the posterior ensemble, 91.3\% of symbolic draws rise continuously toward equal mass, and only 14.7\% show any mathematical evidence of an interior peak. \\

Taken together, these symbolic results isolate a critical new insight: the conditional mass-ratio structure in GWTC-4 is not best interpreted as evidence for two sharply distinct pairing mechanisms. Rather, SR analytically demonstrates that both mass regimes share a fundamental, structural preference for near-equal-mass binaries ($q \to 1$). The primary departure between the two populations is not qualitative, but manifests specifically in how strongly the unequal-mass tails are suppressed (as isolated by the double-exponential cutoff in the low-mass equation). This finding is astrophysically significant for assessing binary formation channels. It implies that the underlying physical pairing mechanism is likely the same across the mass spectrum, e.g., both strongly driven by common-envelope evolution or isolated binary stability \citep{MandelDeMink2016, Bavera2022}. The structural differences isolated by SR likely reflect modest, mass-dependent shifts in the efficiency of these shared processes, such as lower-mass systems being more susceptible to asymmetric supernova kicks disrupting highly unequal-mass pairings \citep{Repetto2012}. \\







\section{Discussion} \label{sec:discussion}

This work has applied symbolic regression to four population relationships from the GWTC-4 BBH catalog, extending \citet{WongCranmer2022} from a single population quantity (the GWTC-3 primary mass spectrum) to merger rates, spin correlations, and mass-dependent pairing, while propagating full posterior uncertainty through draw-by-draw fitting of 200 samples per relation. The analytic gradients enabled by the symbolic surrogates provide a common diagnostic prescription across the fitted relations, and are especially useful for summarizing the flexible models in compact form. \\

The model-independent recovery of $\gamma_0 \approx 3.2$, steeper than the cosmic star formation rate slope of $\sim 2.7$ \citep{MadauDickinson2014}, implies short merger time delays ($t_d \lesssim 1$~Gyr), consistent with isolated binary evolution \citep{MandelDeMink2016, Bavera2022} and dynamical formation in young clusters \citep{Rodriguez2016}. Unlike the parametric $\kappa$ in \citet{GWTC-4_populations}, this slope is recovered without assuming a power-law form. The functional form classification reveals a clear dichotomy: the \textsc{PowerLawRedshift} posterior is overwhelmingly monotonic (194/200 draws), while the \textsc{BSplineIID} model admits a high-redshift turnover with $z_\mathrm{peak} = 1.75^{+0.61}_{-0.92}$, physically expected as the SFR declines \citep{Fishbach2018} but not yet statistically required. Across both the $q$--$\chi_\mathrm{eff}$ and $z$--$\chi_\mathrm{eff}$ analyses, a consistent pattern emerges: the mean effective spin trends are model-dependent and often statistically marginal, whereas the width evolution is robust. The narrowing of $\sigma_{\chi_\mathrm{eff}}(q)$ toward equal mass holds across both \textsc{Linear} ($\mathrm{Pr}[\text{narrows}] = 0.98$) and \textsc{Spline} ($0.81$) models, and the broadening of $\sigma_{\chi_\mathrm{eff}}$ with redshift is confirmed with $\mathrm{Pr}[\text{broadens}] = 1.0$ (\textsc{Linear}) and $0.93$ (\textsc{Spline}). This broadening is consistent with an increasing contribution from dynamical formation channels at higher redshift \citep{Farr2017, Zevin2021}, which produce isotropic spins compared to the aligned distributions expected from field binaries \citep{MandelDeMink2016, Bavera2022}. The SR gradient analysis quantitatively demonstrates that the population's spin evolution is driven primarily by dispersion growth rather than a shift in the mean. \\

The conditional mass-ratio distributions for the two mass peaks both favor near-equal-mass pairing, but the symbolic expressions reveal that the key difference lies in the severity of low-$q$ suppression: a sharp double-exponential cutoff for the 10~$M_\odot$ peak versus a smoother logarithmic decline for the 35~$M_\odot$ peak. This may reflect mass-dependent variations in the efficiency of shared formation processes, such as greater susceptibility of lower-mass systems to natal kicks \citep{Repetto2012}. The compact formulae derived here are directly portable to downstream calculations including rate forecasting, stochastic background estimation, and formation channel comparison, without requiring interpolation of posterior grids. The draw-by-draw ensemble approach mitigates the inherent stochasticity of SR by providing distributions over functional forms rather than single expressions. As the BBH catalog grows, this framework can be reapplied to track how inferred population laws evolve and which features sharpen into robust constraints. \\

\section{Code and Data Availability}

All codes for reproducing the analysis are available on this GitHub \href{https://github.com/chayanchatterjee/Symbolic-Regression-GWTC-4?tab=readme-ov-file}{repo}. The data used in this analysis was obtained from the GWTC-4.0: Population Properties of Merging Compact Binaries public release on  \href{https://zenodo.org/records/16911563}{Zenodo}.

\begin{acknowledgments}
The author would like to thank Cailin Plunkett, Ignacio Magana Hernandez, Thomas Callister and Karan Jani for helpful comments and suggestions. This research was undertaken with the support of compute grant and resources located at Vanderbilt University, USA. This material is based upon work supported by NSF's LIGO Laboratory which is a major facility fully funded by the National Science Foundation. This research used data obtained from the Gravitational Wave Open Science Center (https://www.gw-openscience.org), a service of LIGO Laboratory, the LIGO Scientific Collaboration and the Virgo Collaboration. LIGO is funded by the U.S. National Science Foundation. Virgo is funded by the French Centre National de Recherche Scientifique (CNRS), the Italian Istituto Nazionale della Fisica Nucleare (INFN) and the Dutch Nikhef, with contributions by Polish and Hungarian institutes.
\end{acknowledgments}


\bibliography{references}{}
\bibliographystyle{aasjournalv7}

\end{document}